\documentclass[twocolumn,a4paper,showpacs,prl,aps]{revtex4-1}

\usepackage[dvipsnames]{xcolor}
 \usepackage{amsmath}
 \usepackage{mathrsfs}
 \usepackage{amssymb}
 \usepackage{bbold}
 \usepackage{latexsym}
 \usepackage{amsfonts}
 \usepackage[caption=false]{subfig}
 \usepackage{epsfig}
 \usepackage{psfrag}
 \usepackage{color}
 \definecolor{darkblue}{rgb}{0,0,.5}
 \usepackage[linktocpage, colorlinks=true ,linkcolor=darkblue, citecolor=darkblue]{hyperref}
 \usepackage[all]{hypcap}
 
\usepackage{graphicx}
\usepackage{dcolumn}
\usepackage{bbm}
\usepackage{bm}
\usepackage{overpic}

  \newcommand{\ket}[1]{\left|#1\right>}
  \newcommand{\bra}[1]{\left<#1\right|}
  \newcommand{\expval}[1]{\left< #1 \right>}

\begin{document}

\title{Quantum signatures of Chimera states}
%
%
\author{V. M. Bastidas$^1$}\email{victor@physik.tu-berlin.de}
\author{I. Omelchenko$^1$}
\author{A. Zakharova$^1$}
\author{E. Sch\"oll$^1$}
\author{T. Brandes$^1$}

\affiliation{$^1$Institut f\"ur Theoretische Physik, Technische Universit\"at 
Berlin, Hardenbergstr. 36, 10623 Berlin, Germany}

\begin{abstract}
Chimera states are complex spatiotemporal patterns in networks of identical oscillators, characterized by the coexistence of synchronized and desynchronized dynamics.
Here we propose to extend the phenomenon of chimera states to the quantum regime, and uncover intriguing quantum signatures of these states.
We calculate the quantum fluctuations about semiclassical trajectories and demonstrate that chimera states in the quantum regime can be characterized by bosonic squeezing, weighted quantum correlations, and measures of mutual information. Our findings reveal the relation of chimera states to quantum information theory, and give promising directions for experimental realization of chimera states in quantum systems.

\end{abstract}
%

\maketitle
In classical systems of coupled nonlinear oscillators, the phenomenon of chimera states, which describes the spontaneous emergence
of coexisting synchronized and desynchronized dynamics in networks of identical elements, has recently aroused much interest~\cite{PAN15}. These intriguing spatio-temporal patterns were originally discovered in models of coupled phase oscillators. In this case, they exist due to nonlocal coupling between identical elements of the ensemble~\cite{Kuramoto,Abrams}. There has been extensive work on the theoretical investigation of chimera states~\cite{Atay,Laing,Motter,MartenStrogatz,Torcini,Rosenblum,Lakshmanan, OmelchenkoWolfrum,OmelchenkoSchoell,OmelchenkoHoevel,Zakharova,ASH15,OME15a}, followed by spectacular experimental observations of chimera states in optical~\cite{HagerstromSchoell}, chemical~\cite{Showalter}, mechanical~\cite{MartensHallatschek,KAP14}, electronic~\cite{Maistrenko,ROS14a}, optoelectronic \cite{LAR14}, and electrochemical~\cite{WIC13,Schmidt} setups.

While synchronization of classical oscillators has been well studied since the early observations of Huygens in the $17$th century~\cite{Kapitaniak}, synchronization in quantum mechanics has only very recently become a focus of interest. For example, quantum signatures of synchronization in a network of globally coupled Van der Pol oscillators  have been investigated~\cite{Sadeghpour,Bruder}. Related works focus on the dynamical phase transitions of a network of nanomechanical oscillators with arbitrary topologies characterized by a coordination number~\cite{MarquadOptMech}, and the semiclassical quantization of the Kuramoto model by using path integral methods~\cite{PachonKuramoto}.  

Contrary to classical mechanics, in quantum mechanics the notion of phase-space trajectory is not well defined. As a consequence, one has to define new measures of synchronization for continuous variable systems like optomechanical arrays~\cite{MarquadOptMech}. These measures are based on quadratures of the coupled systems and allow one to extend the notion of phase synchronization to the quantum regime~\cite{Fazio}. Additional measures of synchronization open intriguing connections to concepts of quantum information theory~\cite{GaussianQuantumInformation}, such as decoherence-free subspaces~\cite{Zambrini}, quantum discord~\cite{ZambriniOsc},  entanglement~\cite{LeeWang, ZambriniSpin}, and mutual information~\cite{FazioMutInfo}.
Despite the intensive theoretical investigation of quantum signatures of synchronized states, to date, studies of the quantum manifestations of chimera states are still lacking.

In this Letter we study the emergence of chimera states in a network of coupled quantum Van der Pol oscillators. Unlike in previous work~\cite{Viennot}, we address here the fundamental issue of the dynamical properties of chimera states in a continuous variable system. Considering the chaotic nature of chimera states~\cite{OmelchenkoWolfrum}, we study the short-time evolution of the quantum fluctuations at the Gaussian level. This approach allows us to use powerful tools of quantum information theory to describe the correlations in a nonequilibrium state of the system. We show that quantum manifestations of the chimera state appear in the covariance matrix and are related to bosonic squeezing, thus bringing these signatures into the realm of observability in trapped ions~\cite{Sadeghpour}, optomechanical arrays~\cite{MarquadOptMech}, and driven-dissipative Bose-Einstein condensates~\cite{Diehl}. We find that the chimera states can be characterized in terms of R{\'e}nyi  quantum mutual 
information. Our results reveal that the mutual information for a chimera state lies between the values for synchronized and desynchronized states, which extends in a natural way the definition of chimera states to quantum mechanics.

\paragraph{The model.---}
Similarly to Ref.~\cite{Sadeghpour}, we consider a quantum network consisting of a ring of $N$ coupled Van der Pol oscillators.
Such a network can be described by the master equation for the density matrix $\rho(t)$~\cite{Carmichael}
\begin{align}
      \label{MasterEqNetwork}
            \dot{\rho}&=-\frac{\mathrm{i}}{\hbar}[\hat{H},\rho]+2\sum^{N}_{l=1}\left[\kappa_{1}\mathcal{D}(a_{l}^{\dagger})+\kappa_{2}\mathcal{D}(a_{l}^{2})\right]
      \ ,
\end{align}
where $a^{\dagger}_{l},a_{l}$ are creation and annihilation operators of bosonic particles and  $\mathcal{D}(\hat{O})=\hat{O}\rho\hat{O}^{\dagger}-\frac{1}{2}(\hat{O}^{\dagger}
\hat{O}\rho+\rho\hat{O}^{\dagger}
\hat{O})$ describes dissipative processes with rates $\kappa_1,\kappa_{2}>0$.
In addition, we have imposed periodic boundary conditions $a_{l}=a_{l+N}$ for the bosonic operators.
In contrast to Ref.~\cite{Sadeghpour}, we consider a nonlocal coupling between the oscillators.
Therefore, the Hamiltonian in the interaction picture reads $\hat{H}=\frac{\hbar V}{2d}\sum_{l=1}^{N}\sum_{m=l-d}^{l+d}(a^{\dagger}_{l}a_{m}+a_{l}a^{\dagger}_{m})$,
where $V$ is the coupling strength, $d$ is the coupling range, and $m\neq l$ in the second sum. This kind of coupling implies that  Eq.~\eqref{MasterEqNetwork} has a rotational $S^1$ symmetry. 
In the particular case $d=(N+1)/2$, $N$ odd, one has all-to-all coupling and recovers the results of Ref.~\cite{Sadeghpour}. 

\paragraph{Gaussian quantum fluctuations about semiclassical trajectories .---}
Our aim in this section is to discuss the quantum fluctuations about a semiclassical trajectory in a way similar to Ref.~\cite{Clerk}. We define the expansion of the bosonic operator $a_{l}(t)=\alpha_{l}(t)+\tilde{a}_{l}$, where $\bm{\alpha}(t)=[\alpha_1(t),\ldots,\alpha_{N}(t)]$ is the semiclassical trajectory, and $\hat{\bm{\tilde{a}}}=(\tilde{a}_{1},\ldots,\tilde{a}_{N})$ describes the quantum fluctuations.
In this work we consider the semiclassical regime, where the magnitude of $\alpha_{l}(t)$ is much larger than the quantum fluctuations described by $\tilde{a}_{l}$. 

Let us consider the displacement operator $\hat{D}\left[\bm{\alpha}(t)\right]=\exp\left[\bm{\alpha}(t) \cdot \hat{\bm{\tilde{a}}}^{\dagger}-\bm{\alpha}^{*}(t)\cdot\hat{\bm{\tilde{a}}}\right]$, which enables us to define coherent states $\ket{\alpha_{l}(t)}=\hat{D}\left[\bm{\alpha}(t)\right]\ket{0_{l}}$, where $\ket{0_{l}}$ is the vacuum state of the $l$-th oscillator, and $a_l(t)\ket{\alpha_{l}(t)}=\alpha_{l}(t)\ket{\alpha_{l}(t)}$~\cite{Glauber}. By using the expansion of the master equation about the mean-field  $\bm{\alpha}(t)$ described in the supplemental material~\cite{SI}, we obtain a master equation for the  density operator in a co-moving frame $\rho_{\bm{\alpha}}(t)=\hat{D}^{\dagger}\left[\bm{\alpha}(t)\right]\rho(t)\hat{D}\left[\bm{\alpha}(t)\right]$ 
\begin{equation}
      \label{SemiclassMasterNetwork}
        \dot{\rho}_{\bm{\alpha}}\approx-\frac{\mathrm{i}}{\hbar}[\hat{H}_{\text{Q}}^{(\bm{\alpha})},\rho_{\bm{\alpha}}]+2\sum^{N}_{l=1}\left[\kappa_{1}\mathcal{D}(\tilde{a}_{l}^{\dagger})
        +4\kappa_{2}|\alpha_{l}|^2\mathcal{D}(\tilde{a}_{l})\right]
      \ .
\end{equation}
In addition, the coherent dynamics of the fluctuations is governed by
the Hamiltonian
\begin{equation}
      \label{EffQuadHamNet}
            H^{(\bm{\alpha})}_{\text{Q}}=\mathrm{i}\hbar\sum^{N}_{l=1}\kappa_{2}(\alpha_{l}^{*})^{2}\tilde{a}_{l}^{2}
            +\frac{\hbar V}{2d}\sum^{N}_{l=1}\sum^{l+d}_{\substack{m=l-d\\m\neq l}}\tilde{a}^{\dagger}_{l}\tilde{a}_{m}+\text{H.c}
       \ .
\end{equation}

\begin{figure}
\includegraphics[width=0.49\textwidth,clip=true]{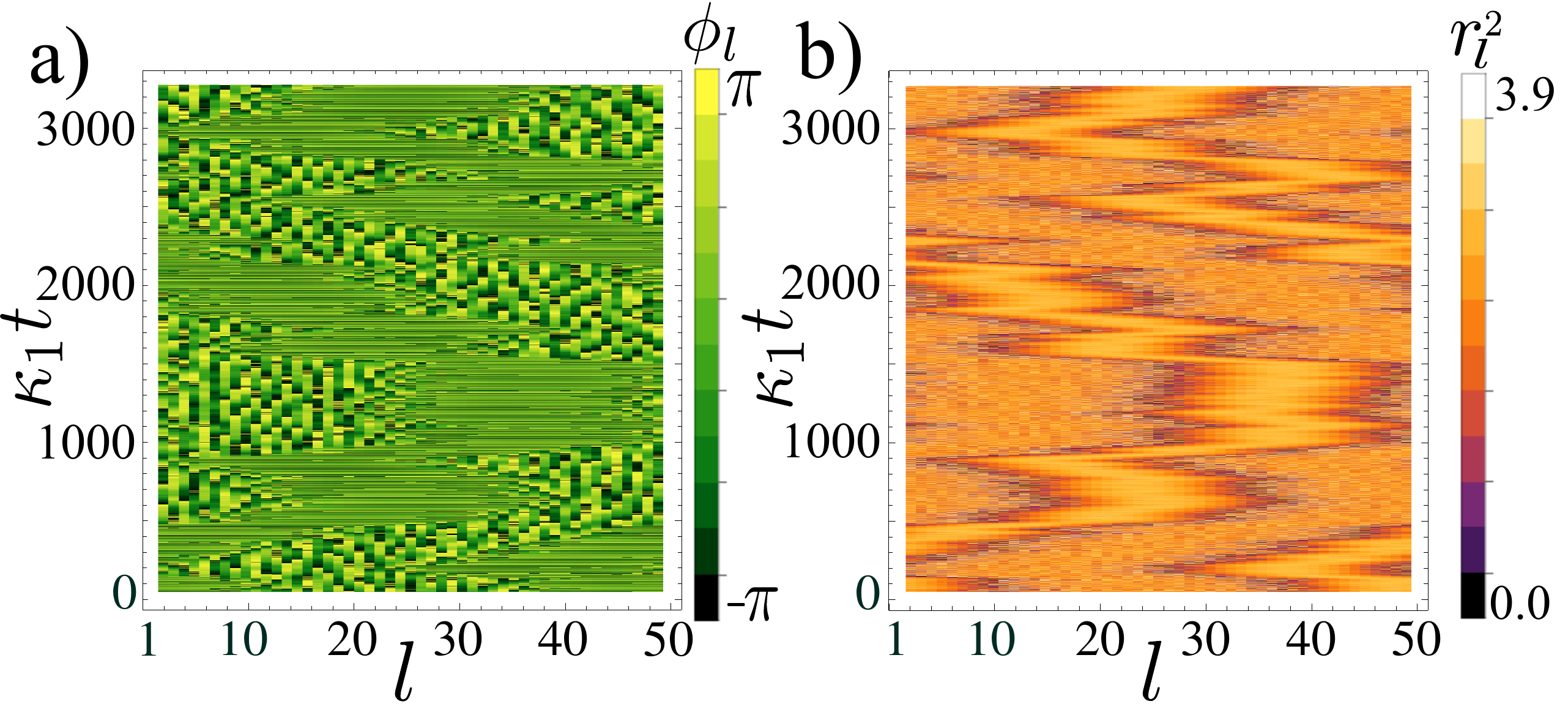}
    \caption{
    \label{Fig1}
   Space-time representation of the classical chimera state for oscillators
   $\alpha_{l}(t)=r_{l}(t)e^{\mathrm{i}\phi_{l}(t)}$: (a) $\phi_{l}(t)$ and (b) $r^{2}_{l}(t)$.  Parameters: $d=10$, $\kappa_2=0.2 \kappa_1$, $V=1.2\kappa_1$, and $N=50$.}
\end{figure}
The mean fields appearing in Eq.~\eqref{EffQuadHamNet} satisfy the equation of motion
\begin{equation}
      \label{QuantEqMotionNetwork}
            \dot{\alpha}_{l}(t)=\alpha_{l}(t)(\kappa_{1}-2\kappa_{2}|\alpha_{l}(t)|^2)-\mathrm{i}\frac{V}{2d}\sum^{l+d}_{\substack{m=l-d\\m\neq l}}\alpha_{m}(t)
\end{equation}
with a similar equation for $\dot{\alpha}^{*}_{l}(t)$. The equations of motion Eq.~\eqref{QuantEqMotionNetwork} resemble a system of coupled Stuart-Landau oscillators~\cite{Zakharova}.
By solving the equations of motion Eq.~\eqref{QuantEqMotionNetwork}, one obtains the time-dependent mean field $\bm{\alpha}(t)$. Such a mean field plays a fundamental role in the description of the master equation  Eq.~\eqref{SemiclassMasterNetwork}. In particular, the mean field drives coherent effects such as squeezing in Eq.~\eqref{EffQuadHamNet} and it determines the time-dependent rates which appear in Eq.~\eqref{SemiclassMasterNetwork}.

\begin{figure}
\includegraphics[width=0.49\textwidth,clip=true]{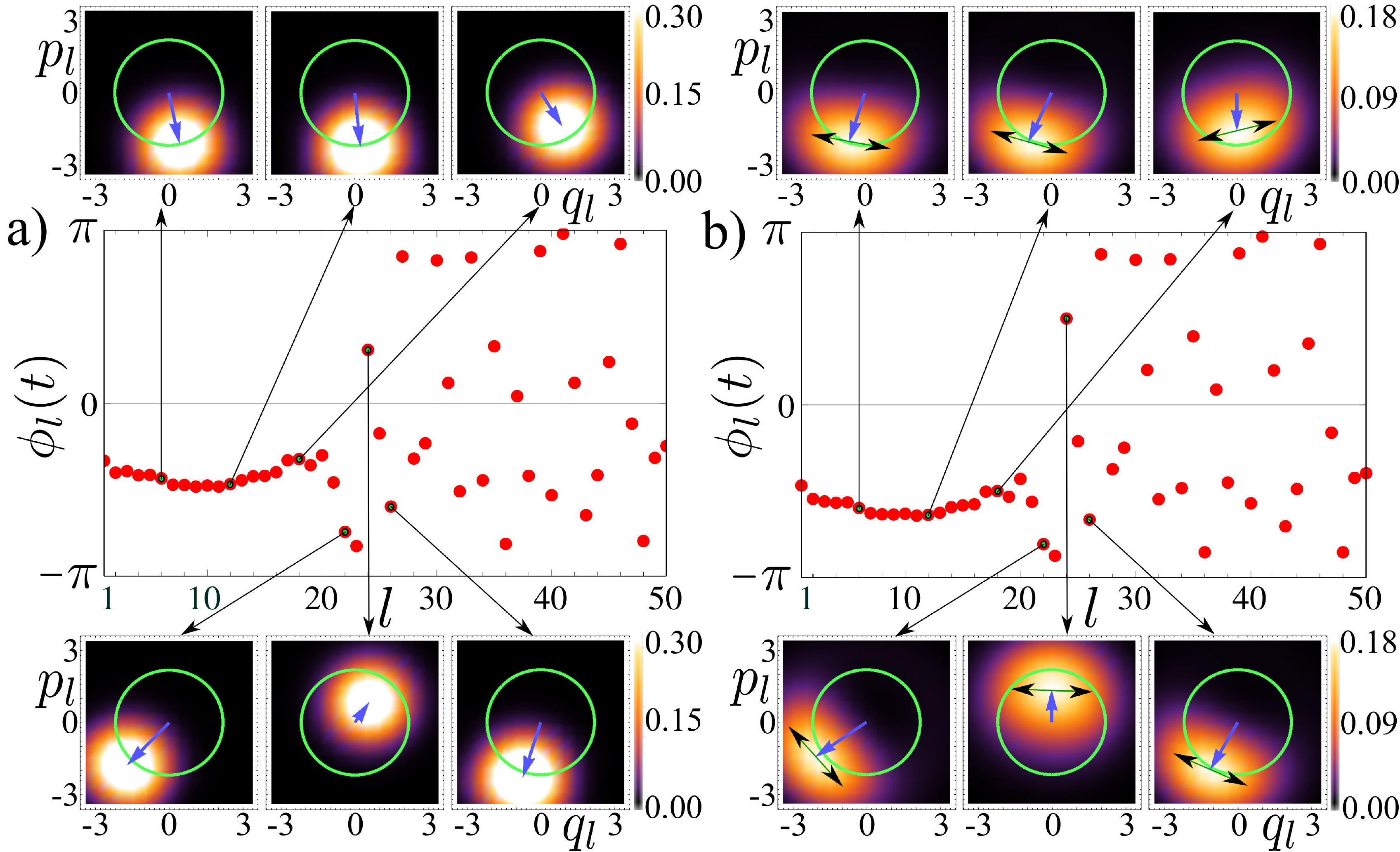}
    \caption{
    \label{Fig2}
   Quantum signatures of the classical chimera state.  (a) Snapshot of the phase chimera depicted in Fig.~\ref{Fig1} at $\kappa_{1}t_{0}=3000$. We consider an initial density matrix $\rho(t_{0})$ which is a tensor product of coherent states centered around the positions of the individual oscillators as depicted in the insets (Husimi function). (b)
   After a short-time interval $\kappa_{1}\Delta t=0.5$, quantum correlations appear in the form of squeezing (black double arrows in the insets). Parameters: $d=10$, $\kappa_2=0.2 \kappa_1$, $V=1.2\kappa_1$, and $N=50$.}
\end{figure}

\paragraph{The Classical Chimera state.---}
From our previous discussion, the classical equations of motion~\eqref{QuantEqMotionNetwork} must be satisfied in order to describe the physics in the co-moving frame. In the polar representation $\alpha_{l}(t)=r_{l}(t)e^{\mathrm{i}\phi_{l}(t)}$
the equations of motion couple amplitude $r_{l}(t)$ and phase $\phi_{l}(t)$ of the individual oscillators. 
We numerically solve Eq.~\eqref{QuantEqMotionNetwork} for a network of $N=50$ coupled oscillators with coupling range $d=10$, considering initial conditions $|\alpha_{l}(t_{0})|\approx r_{0}$, where $r_{0}=1.58$, and phases drawn randomly from a Gaussian distribution in space~\cite{SI}. Figure~\ref{Fig1} depicts the time evolution of a classical chimera state. In Fig.~\ref{Fig1}~(a) we show the space-time representation of the phases $\phi_{l}(t)$ of the individual oscillators. One can observe that for a fixed time, there is a domain of
synchronized oscillators that coexists with a domain of desynchronized motion, which is a typical feature of chimera states. Besides the phase, also the amplitude exhibits chimera dynamics as we show in Fig.~\ref{Fig1}~(b). 
One can observe that the width of the synchronized region changes with time. Similarly, the center of mass of the synchronized region moves randomly along the ring~\cite{OmelchenkoWolfrum}. In the case of the uncoupled system with $V=0$, the individual oscillators exhibit a limit cycle with radius $r_{0}=\sqrt{\frac{\kappa_{1}}{2\kappa_{2}}}$, which is depicted in the insets of Fig.~\ref{Fig2} by the green circle.

\paragraph{Gaussian fluctuations and the Wigner function.---}
As discussed, the classical equations of motion Eq.~\eqref{QuantEqMotionNetwork} exhibit a chimera state. By using the knowledge we have about the classical trajectory $\bm{\alpha}(t)$, we can study the quantum fluctuations in the co-moving frame by solving the master equation Eq.~\eqref{SemiclassMasterNetwork}. For this purpose, we consider the pure coherent state as an initial density matrix $\rho(t_{0})=\bigotimes_{l=1}^{N}\ket{\alpha_{l}(t_{0})}\bra{\alpha_{l}(t_{0})}$, where $|\alpha_{l}(t_{0})|\approx 1.58$ and we choose the phases as in the left panel of Fig.~\ref{Fig2}. This initial condition corresponds to a fixed time $t_0=3000/\kappa_{1}$ in Fig.~\ref{Fig1}. In the co-moving frame, such initial condition reads $\rho_{\bm{\alpha}}(t_{0})=\bigotimes_{l=1}^{N}\ket{0_{l}}\bra{0_{l}}$.
For convenience, let us write the bosonic operators $a_{l}=(\hat{q}_{l}+\mathrm{i}\hat{p}_{l})/\sqrt{2\hbar}$ and $\tilde{a}_{l}=(\hat{\tilde{q}}_{l}+\mathrm{i}\hat{\tilde{p}}_{l})/\sqrt{2\hbar}$ in terms of position and momentum operators. In terms of complex variables $z_{l}=(q_{l}+\mathrm{i}p_{l})/\sqrt{2\hbar}$, $\tilde{z}_{l}=(\tilde{q}_{l}+\mathrm{i}\tilde{p}_{l})/\sqrt{2\hbar}$, we define the coordinates $\bm{z}^{T}=(z_{1},\ldots,z_{N})$ in the laboratory frame and $\bm{\tilde{z}}^{T}=(\tilde{z}_{1},\ldots,\tilde{z}_{N})$ in the co-moving frame, such that $\bm{z}=\bm{\alpha}(t)+\bm{\tilde{z}}$. The variables $q_{l},\tilde{q}_{l}$ and $p_{l},\tilde{p}_{l}$ denote position and conjugate momentum, respectively.

We consider the Wigner representation $W_{\bm{\alpha}}(\bm{\tilde{R}},t)$ of the density operator $\rho_{\bm{\alpha}}(t)$, where $\bm{\tilde{R}}^{T}=(\tilde{q}_{1},\tilde{p}_{1},\dots,\tilde{q}_{N},\tilde{p}_{N})$. By using standard techniques of quantum optics~\cite{Carmichael}, the master equation Eq.~\eqref{SemiclassMasterNetwork} can be represented as a Fokker-Planck equation for the Wigner function which depends on the mean field solution of Eq.~\eqref{QuantEqMotionNetwork} and contains information of the chimera state.
In the supplemental material~\cite{SI}, we provide the explicit form of the Fokker-Planck equation for $W_{\bm{\alpha}}$.
Fortunately, even though the coefficients of the equation are time dependent, one can derive an exact solution $W_{\bm{\alpha}}(\bm{\tilde{R}},t)=(2\pi)^{-N}(\det \mathscr{C})^{-1/2}\exp\left(-\frac{1}{2}\bm{\tilde{R}}^{T}\cdot\mathscr{C}^{-1}\cdot\bm{\tilde{R}}\right)$,
where $\mathscr{C}(t)$ is the covariance matrix, whose matrix elements $\mathscr{C}_{ij}=\expval{\frac{1}{2}(\hat{\tilde{R}}_{i}\hat{\tilde{R}}_{j}+\hat{\tilde{R}}_{j}\hat{\tilde{R}}_{i})}_{\bm{\alpha}
}-\expval{\hat{\tilde{R}}_{i}}_{\bm{\alpha}}\expval{\hat{\tilde{R}}_{j}}_{\bm{\alpha}}$ include information about the correlations between quantum fluctuations $\hat{\tilde{R}}_{2l-1}=\hat{\tilde{q}}_{l}$ and $\hat{\tilde{R}}_{2l}=\hat{\tilde{p}}_{l}$. The angular brackets $\expval{\hat{O}}_{\bm{\alpha}}=\text{tr}(\rho_{\bm{\alpha}}\hat{O})$ denote the expectation value of an operator $\hat{O}$ calculated with the density matrix $\rho_{\bm{\alpha}}$.

The solution $W_{\bm{\alpha}}(\bm{\tilde{R}},t)$ corresponds to a Gaussian distribution centered at the origin in the co-moving frame. In the laboratory frame, the Wigner function is centered at the classical trajectory $\bm{\alpha}(t)$. However, due to the chaotic nature of the classical chimera state~\cite{OmelchenkoWolfrum}, our exact solution is just valid for short-time evolution. The Husimi function $Q(\bm{z})=\frac{1}{\pi}\bra{\bm{z}}\rho(t)\ket{\bm{z}}$ is intimately related to the Wigner function~\cite{Carmichael, SI} and can be obtained numerically by using the  Gutzwiller ansatz~\cite{Sadeghpour}. The insets in the right panel of Fig.~\ref{Fig2} depict the Husimi functions of the individual nodes after a short evolution time $\Delta t=0.5/\kappa_1$. 
One can observe that even if one prepares the system in a separable state, quantum fluctuations arise in the form of bosonic squeezing of the oscillators~\cite{Carmichael}. In the insets of Fig.~\ref{Fig2}, the arrows indicate the direction perpendicular to the squeezing direction for the individual oscillators. For oscillators within the synchronized region, the squeezing occurs almost in the same direction. In contrast, the direction of squeezing is random for oscillators in the desynchronized region, which reflects the nature of the chimera state.
\paragraph{Quantum signatures of a chimera state in the covariance matrix.---}
Now let us study the consequences of the exact solution for the short-time evolution of the Wigner function. Once we obtain the solution of the equations of motion Eq.~\eqref{QuantEqMotionNetwork}, one can find the corresponding covariance matrix $\mathscr{C}(t)$. As we have defined in the introduction, a chimera state is characterized by the coexistence in space of synchronized and desynchronized motion. Therefore, to understand the quantum manifestations of a chimera state, we require to study also quantum signatures of synchronized and desynchronized dynamics.

\begin{figure}
\includegraphics[width=0.49\textwidth,clip=true]{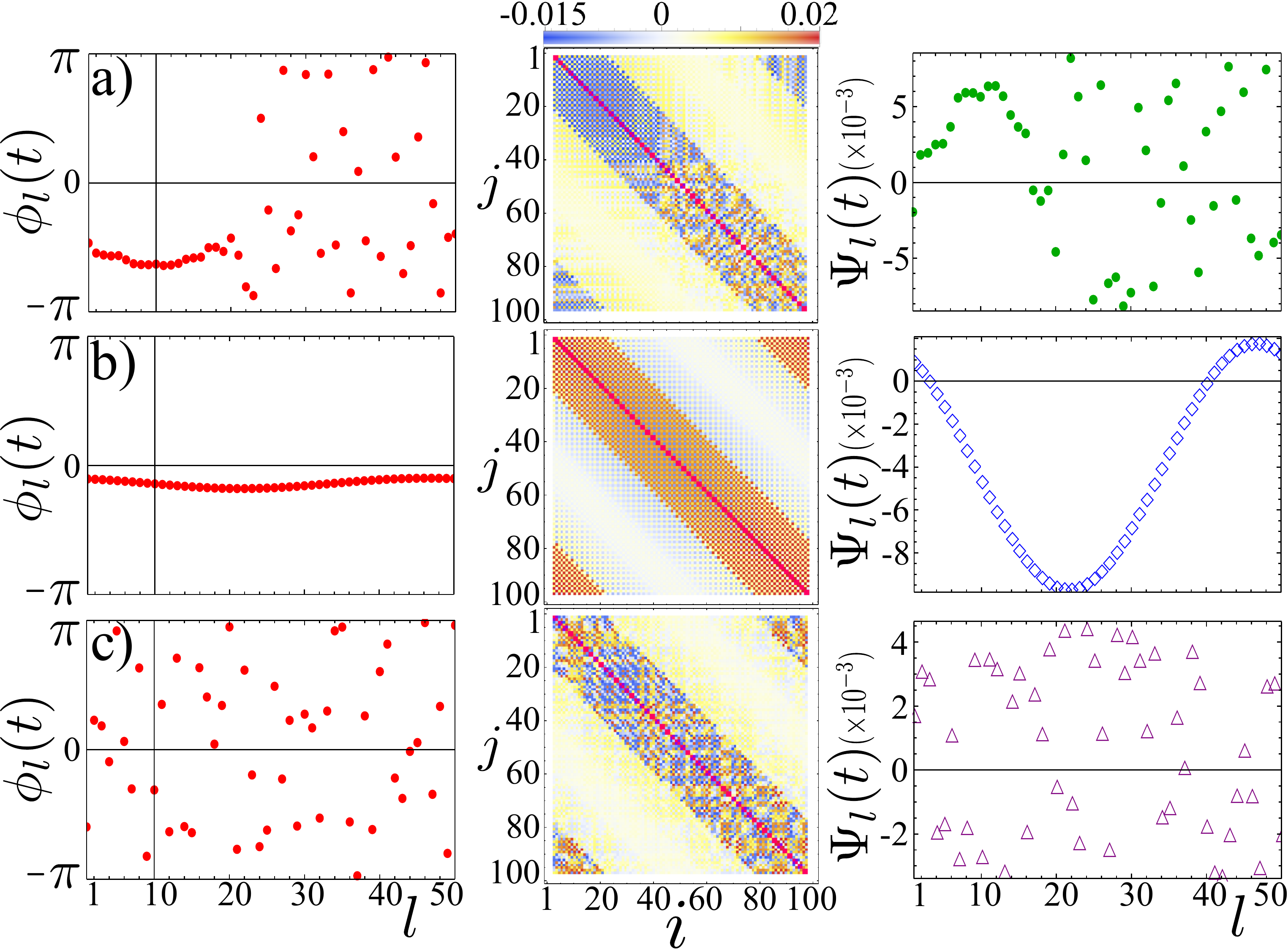}
    \caption{
    \label{Fig3}
   Quantum fluctuations after a short-time evolution. Similarly to Fig.~\ref{Fig2}, we consider an initial density matrix $\rho(t_{i})$ which is a tensor product of coherent states centered around the classical positions of the oscillators.
   Snapshots of the phase (left column) and covariance matrices (central column) after short-time evolution $\kappa_{1}\Delta t=0.5$ of the states: (a) chimera for $V=1.2\kappa_{1}$, (b) synchronized state for $V=1.6\kappa_{1}$, and (c) desynchronized state for $V=0.8\kappa_{1}$. Right column: Weighted spatial average $\Psi_{l}(t)$ of the covariance matrix for the states shown in a), b) and c), respectively. Parameters $d=10$, $\kappa_2=0.2 \kappa_1$, and $N=50$. 
   }
\end{figure}

The left column of Fig.~\ref{Fig3} show snapshots of the phases for (a) chimera, (b) synchronized, and (c) desynchronized mean-field solutions of Eq.~\eqref{QuantEqMotionNetwork}. The central column of Fig.~\ref{Fig3} depicts the corresponding covariance matrices after a short evolution time $\Delta t=0.5/\kappa_{1}$.
For every plot, we have initialized the system at time $t_{i}$ as a tensor product of coherent states $\ket{\alpha_{l}(t_{i})}$ centered at the positions $\alpha_{l}(t_{i})$ of the individual oscillators. As a consequence, the covariance matrix at the initial time is diagonal $\mathscr{C}_{2l-1, 2l-1}(t_i)=\expval{\hat{\tilde{q}}^{2}_{l}}_{\bm{\alpha}}=\hbar/2$ and $\mathscr{C}_{2l, 2l}(t_i)=\expval{\hat{\tilde{p}}^{2}_{l}}_{\bm{\alpha}}=\hbar/2$, which reflects the Heisenberg uncertainty principle because $\expval{\hat{\tilde{q}}_{l}}_{\bm{\alpha}}=\expval{\hat{\tilde{p}}_{l}}_{\bm{\alpha}}=0$.

After a short evolution time, quantum correlations are built up due to the coupling between the oscillators, and the covariance matrix exhibits a nontrivial structure which is influenced by the mean field solution. For example, the central panel of Fig.~\ref{Fig3}~a) shows a matrix plot of the covariance matrix corresponding to a chimera state obtained from the same initial condition as in Fig.~\eqref{Fig2}. The covariance matrix acquires a block structure, where the upper $40\times 40 $ block (corresponding to nodes $l=1,\dots,20$) shows a regular pattern matching the synchronized region of the chimera state. Similarly, the lower $60\times 60$ block shows an irregular structure which corresponds to the desynchronized dynamics of the oscillators $l=21,\dots,50$. In a similar fashion, Figs.~\ref{Fig3}~b) and c) show the matrix $\mathscr{C}$ for completely synchronized and desynchronized states, respectively.
In the case of a chimera state, this coincides with the results shown in Fig.~\ref{Fig2}, where the squeezing direction of the oscillators is related to the classical solution. In order to quantify these observations we define the weighted  correlation as
\begin{equation}
      \label{KuramotoOrderPar}
            \Psi_{l}(t)=\frac{V}{2d}\sum^{l+d}_{\substack{m=l-d\\m\neq l}}\mathscr{C}_{2l, 2m}(t)
      \ .
\end{equation}
This spatial average 
highlights the structure of the covariance matrix. The right column of Fig.~\ref{Fig3} shows 
$\Psi_{l}(t)$ for a) chimera, b) synchronized, and c) desynchronized states. The chimera state exhibits a regular and an irregular domain, exactly as the classical chimera does. 

\paragraph{Chimera states and R{\'e}nyi quantum mutual information.---}
Now let us consider a partition of the network into spatial domains of size $L$  and $N-L$, which we call \textit{Alice} (A) and \textit{Bob} (B), respectively. This partition can be represented by considering a decomposition of the covariance matrix
\begin{equation}
       \label{CovMatrix}
       \mathscr{C}(t) =\left(%
\begin{array}{ccc}
\mathscr{C}_{\text{A}}(t)  &  \mathscr{C}_{\text{AB}}(t) \\
 \mathscr{C}^{T}_{\text{AB}}(t) & \mathscr{C}_{\text{B}}(t)
\end{array}
\right)
\end{equation}
\begin{figure}
\includegraphics[width=0.46\textwidth,clip=true]{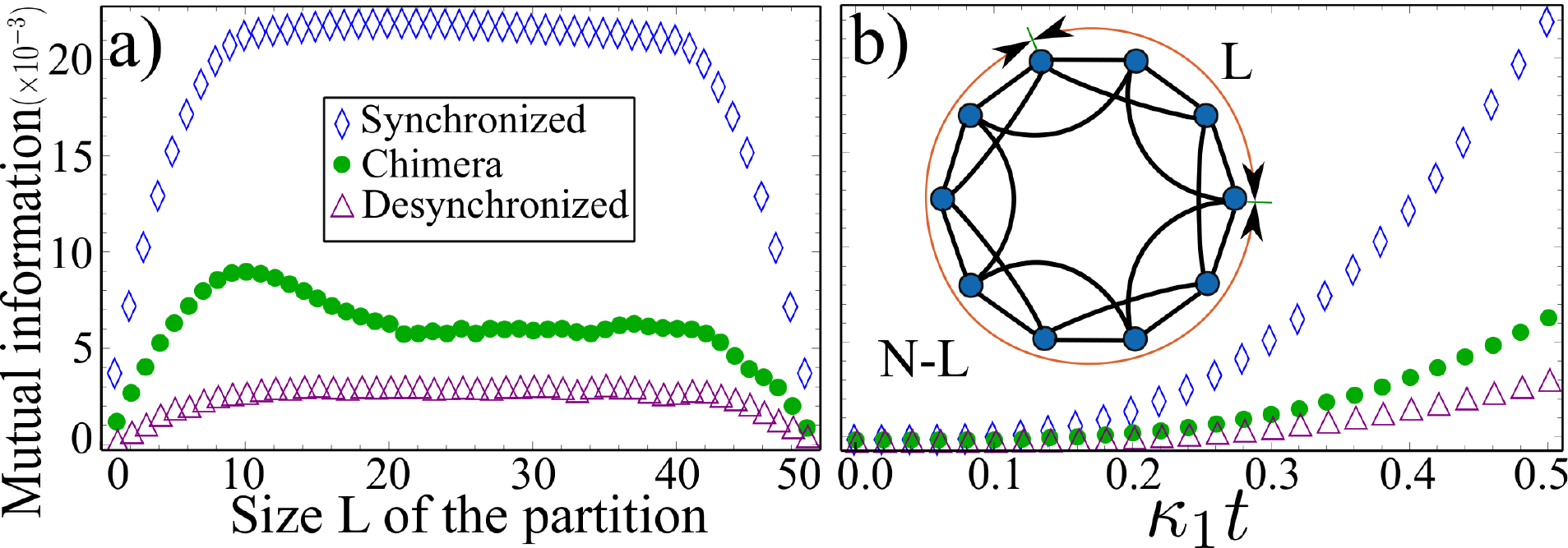}
    \caption{
    \label{Fig4}
   R{\'e}nyi quantum mutual information for the states shown in Fig.~\ref{Fig3}. The green dots, blue diamonds, and purple triangles represent the chimera, synchronized, and desynchronized states, respectively. a) Gaussian R{\'e}nyi-2  mutual information $\mathcal{I}_{2}(\rho_{\text{A:B}})$ as a function of the size $L$ of \textit{Alice} after an evolution time $\Delta t=0.5/\kappa_{1}$. b) The time evolution of the mutual information during the time interval $\Delta t$ for a fixed size $L_{c}=20$. Inset: scheme of the nonlocally coupled network. 
   Parameters: $d=10$, $\kappa_2=0.2\kappa_1$, and $N=50$.}
\end{figure}
To study the  interplay between synchronized and desynchronized dynamics, which is characteristic of a chimera state, we propose the use of an entropy measure~\cite{GaussianQuantumInformation,FazioMutInfo}. Of particular interest is the R{\'e}nyi entropy $S_{\mu}(\rho)=(1-\mu)^{-1}\ln\text{tr}(\rho^{\mu})$, $\mu \in \mathbb{N}$, of the density matrix $\rho$, which is discussed in Ref.~\cite{Adesso}. In terms of the Wigner representation of $\rho_{\bm{\alpha}}$, the R{\'e}nyi entropy for $\mu=2$ reads $S_{2}(\rho_{\bm{\alpha}})=-\ln\left[\int W^{2}_{\bm{\alpha}}(\bm{\tilde{R}},t) d^{2N}\bm{\tilde{R}} \right]$. Now let us consider the bipartite Gaussian state $\rho_{\text{AB}}=\rho_{\bm{\alpha}}$ composed of \textit{Alice}
and  \textit{Bob} and define the tensor product $\rho_{\text{Ref}}=\rho_{\text{A}}\otimes\rho_{\text{B}}$ of the two marginals $\rho_{\text{A}}$ and $\rho_{\text{B}}$.

To measure Gaussian R{\'e}nyi-2  mutual information $\mathcal{I}_{2}(\rho_{\text{A:B}})=S_{2}(\rho_{\text{A}})+S_{2}(\rho_{\text{B}})-S_{2}(\rho_{\text{AB}})$, we require the calculation of the relative sampling entropy between the total density matrix $\rho_{\text{AB}}$ and the reference state $\rho_{\text{Ref}}$ as shown in Ref.~\cite{Adesso}. This leads to a formula  $\mathcal{I}_{2}(\rho_{\text{A:B}})=\frac{1}{2}\ln\left(\det\mathscr{C}_{A}\det\mathscr{C}_{B}/\det\mathscr{C}\right)$ in terms of the covariance matrix Eq.~\eqref{CovMatrix}.
Figure~\ref{Fig4}~a) shows the variation of $\mathcal{I}_{2}(\rho_{\text{A:B}})$ as a function of the size $L$ of the partition after an evolution time $\Delta t=0.5/\kappa_{1}$. One can observe that for a chimera state the mutual information is asymmetric as a function of $L$ and there is a critical size $L_{c}=20$, where a dramatic change of the correlations occurs.

Now let us consider the chimera state shown in Fig.~\ref{Fig2}, and let us consider a partition where the size of \textit{Alice} is $L_{c}=20$. Fig.~\ref{Fig4}~b) shows the time evolution of mutual information for such a state. In addition, by using the same partition as for the chimera state,
we calculate the mutual information for the synchronized and desynchronized states depicted in Figs.~\ref{Fig3}~b) and c), respectively. Our results reveal that the chimera state has a mutual information which lies between the values for synchronized and desynchronized states. This resembles the definition of a chimera state given at the beginning of the article.

\paragraph{Conclusion.---}
We have shown that quantum signatures of chimera states appear in the squeezing of coherent states, in the covariance matrix, and in measures of mutual information. To quantify the structure of the covariance matrix, we have introduced a spatial average of the quantum correlation, which reveals the nature of the classical trajectory, i.e., chimera, synchronized, or desynchronized state. The mutual information for a bipartite state $\mathcal{I}_{2}(\rho_{\text{A:B}})$ extends the definition of a chimera to the quantum regime and highlights the relation to quantum information theory. A possible experimental realization of our model could be carried out by means of trapped ions~\cite{Wineland}, as it was suggested in Ref.~\cite{Sadeghpour}. Other experimental possibilities include Bose-Einstein condensation in the presence of dissipation and external driving~\cite{Diehl,Kasprzak}. In this context our approach is particularly interesting, because the continuum limit of the mean field 
Eq.~\eqref{QuantEqMotionNetwork} is a complex Ginzburg-Landau 
equation, which is nonlocal in space~\cite{Kuramoto}. In this sense, our linearized master equation Eq.~\eqref{SemiclassMasterNetwork} enables us to study the Bogoliubov excitations around the mean field solution.

\paragraph{Acknowledgments.---}
V.M. Bastidas thanks L. M. Valencia and Y. Sato. The authors acknowledge inspiring discussions with J. Cerrillo, S. Restrepo, G. Schaller, and P. Strasberg. This work was supported by DFG in the framework of SFB 910.

\vspace{-0.7cm}

\newpage
\begin{widetext}
 
\section*{Supplemental material for ``Quantum signatures of Chimera states''}
\begin{center}
      V. M. Bastidas$^1$, I. Omelchenko$^1$, A. Zakharova$^1$, E. Sch\"oll$^1$, and T. Brandes$^1$
\end{center}

$^1$Institut f\"ur Theoretische Physik, Technische Universit\"at 
Berlin, Hardenbergstr. 36, 10623 Berlin, Germany
\section{Calculation of the Gaussian quantum fluctuations for a single Van der Pol oscillator}
In the case of a single Van der Pol oscillator~\cite{SISadeghpour,SIBruder} we can write the master equation Eq.~\eqref{MasterEqNetwork} as follows
\begin{equation}
      \label{SingleVDPME}
            \dot{\rho}(t)=-\frac{\mathrm{i}}{\hbar}[\hat{H}_{\text{int}},\rho]+2\kappa_{1}\left(a^{\dagger}\rho a-\frac{1}{2}\{\rho,a a^{\dagger}\}\right)+2\kappa_{2}\left(a^{2}\rho (a^{\dagger})^{2}-\frac{1}{2}\{\rho,(a^{\dagger})^{2}a^{2}\}\right)
      \ ,
\end{equation}
where in the main text we set $\hat{H}_{\text{int}}=0$ for the single oscillator and only consider coherent dynamics through the coupling.

Following a similar method as the one used in Ref.~\cite{SIClerk}, if one defines the reduced density matrix in the co-moving frame $\rho_{\alpha}(t)=\hat{D}^{\dagger}\left[\alpha(t)\right]\rho(t)\hat{D}\left[\alpha(t)\right]$, we obtain a master equation with Liouville operators $\hat{\mathcal{L}}_{1}$ and $\hat{\mathcal{L}}_{2}$
\begin{align}
      \label{MasterEqRot}
        \dot{\rho}_{\alpha}(t)&=-\frac{\mathrm{i}}{\hbar}[\hat{H}^{(\alpha)}(t),\rho_{\alpha}(t)]+\hat{D}^{\dagger}\left[\alpha(t)\right]\left[2\kappa_{1}\left(\tilde{a}^{\dagger}\rho \tilde{a}-\frac{1}{2}\{\rho,\tilde{a} \tilde{a}^{\dagger}\}\right)+2\kappa_{2}\left(\tilde{a}^{2}\rho (\tilde{a}^{\dagger})^{2}-\frac{1}{2}\{\rho,(\tilde{a}^{\dagger})^{2}\tilde{a}^{2}\}\right)\right]\hat{D}\left[\alpha(t)\right] 
        \nonumber \\&
        \equiv-\frac{\mathrm{i}}{\hbar}[\hat{H}^{(\alpha)}(t),\rho_{\alpha}(t)]+\hat{\mathcal{L}}_{1}\rho_{\alpha}+\hat{\mathcal{L}}_{2}\rho_{\alpha}
      \ ,
\end{align}
where we have defined $\hat{H}^{(\alpha)}(t)=\hat{D}^{\dagger}\left[\alpha(t)\right](\hat{H}_{\text{int}}-\mathrm{i}\hbar\partial_t)\hat{D}\left[\alpha(t)\right]$ and the anticommutator  $\{\hat{A},\hat{B}\}=\hat{A}\hat{B}+\hat{B}\hat{A}$. For later purposes, we need to use the identity
\begin{equation}
      \label{DerDisp}
          \mathrm{i}\hbar\hat{D}^{\dagger}\left[\alpha(t)\right]\partial_t\hat{D}\left[\alpha(t)\right]
          =\frac{\mathrm{i\hbar}}{2}[\dot{\alpha}(t)\alpha^{*}(t)-\alpha(t)\dot{\alpha}^{*}(t)]+\mathrm{i}\hbar[\dot{\alpha}(t)\tilde{a}^{\dagger}-\dot{\alpha}^{*}(t)\tilde{a}]
      \ .
\end{equation}

We are interested in the action of the displacement operator $\hat{D}\left[\alpha(t)\right]$ on the other terms of the master equation. By using elementary properties of the displacement operator we obtain
\begin{align}
      \label{DisplLinDiss}
           \hat{\mathcal{L}}_{1}\rho_{\alpha}(t)&=2\kappa_{1}\hat{D}^{\dagger}\left[\alpha(t)\right]\left(\tilde{a}^{\dagger}\rho \tilde{a}-\frac{1}{2}\{\rho,\tilde{a} \tilde{a}^{\dagger}\}\right)\hat{D}\left[\alpha(t)\right]
           \nonumber\\
           &=2\kappa_{1}\left(\tilde{a}^{\dagger}\rho_{\alpha} \tilde{a}-\frac{1}{2}\{\rho_{\alpha},\tilde{a} \tilde{a}^{\dagger}\}\right)-\frac{\mathrm{i}}{\hbar}\left[\mathrm{i}\hbar\kappa_{1}\alpha \tilde{a}^{\dagger},\rho_{\alpha}\right]-\frac{\mathrm{i}}{\hbar}\left[-\mathrm{i}\hbar\kappa_{1}\alpha^{*}\tilde{a},\rho_{\alpha}\right]
      \ .
\end{align}
However, the dissipative term which is proportional to $\kappa_{2}$ requires more attention. Therefore, we study each term individually
\begin{align}
      \label{DisplNonLinDiss}
           \hat{\mathcal{L}}_{2}\rho_{\alpha}(t)&=2\kappa_{2}\hat{D}^{\dagger}\left[\alpha(t)\right]\left(\tilde{a}^{2}\rho (\tilde{a}^{\dagger})^{2}-\frac{1}{2}\{\rho,(\tilde{a}^{\dagger})^{2}\tilde{a}^{2}\}\right)\hat{D}\left[\alpha(t)\right]
           \nonumber\\
           &=2\kappa_{2}\left(\tilde{a}^{2}\rho_{\alpha} (\tilde{a}^{\dagger})^{2}-\frac{1}{2}\{\rho_{\alpha},(\tilde{a}^{\dagger})^{2}\tilde{a}^{2}\} \right)+4\kappa_{2}\alpha^{*}\left(\tilde{a}^{2}\rho_{\alpha} \tilde{a}^{\dagger}-\frac{1}{2}\{\rho_{\alpha},\tilde{a}^{\dagger}\tilde{a}^{2}\} \right)+4\kappa_{2}\alpha\left(\tilde{a}\rho_{\alpha} (\tilde{a}^{\dagger})^{2}-\frac{1}{2}\{\rho_{\alpha},(\tilde{a}^{\dagger})^{2}\tilde{a}\} \right)
           \nonumber\\&
           +8\kappa_{2}|\alpha|^2\left(\tilde{a}\rho_{\alpha}\tilde{a}^{\dagger}-\frac{1}{2}\{\rho_{\alpha},\tilde{a}^{\dagger}\tilde{a}\} \right)-\frac{\mathrm{i}}{\hbar}\left[\mathrm{i}\hbar\kappa_{2}(\alpha^{*})^{2}\tilde{a}^{2},\rho_{\alpha}\right]-\frac{\mathrm{i}}{\hbar}\left[-\mathrm{i}\hbar\kappa_{2}\alpha^{2}(\tilde{a}^{\dagger})^{2},\rho_{\alpha}\right]
           \nonumber\\&
           -\frac{\mathrm{i}}{\hbar}\left[2\mathrm{i}\hbar\kappa_{2}\alpha (\alpha^{*})^2 \tilde{a},\rho_{\alpha}\right]-\frac{\mathrm{i}}{\hbar}\left[-2\mathrm{i}\hbar\kappa_{2}\alpha^{*}\alpha^2\tilde{a}^{\dagger},\rho_{\alpha}\right]
      \ .
\end{align}
Now we compare with the general form of the Lindblad master equation~\cite{SICarmichael} with Lindblad operators $\hat{L}_{\mu}$
\begin{equation}
      \label{LimbladVDP}
            \dot{\rho}(t)=-\frac{\mathrm{i}}{\hbar}[H_{\text{int}},\rho]+\sum_{\mu}\gamma_{\mu}\left(\hat{L}_{\mu}\rho \hat{L}_{\mu}^{\dagger}-\frac{1}{2}\{\rho,\hat{L}_{\mu}^{\dagger}\hat{L}_{\mu}\}\right)
      \ ,
\end{equation}
which enables us to introduce an effective Hamiltonian to describe the dynamics between quantum jumps
\begin{equation}
      \label{EffGenHam}
            H_{\text{eff}}=H_{\text{int}}-\frac{\mathrm{i}\hbar}{2}\sum_{\mu}\gamma_{\mu}\hat{L}_{\mu}^{\dagger}\hat{L}_{\mu}
      \ .
\end{equation}
In order to study quantum fluctuations about semiclassical trajectories, we consider an effective non-hermitian Hamiltonian as an expansion in powers of quantum fluctuations
\begin{align}
      \label{EffQuadHam}           
            H^{(\alpha)}_{\text{eff}}&=-\mathrm{i}\hbar\kappa_{2}(\tilde{a}^{\dagger})^2\tilde{a}^2+\hbar(-2\mathrm{i}\kappa_{2}\alpha^{*}\tilde{a}^{\dagger}\tilde{a}^{2}-2\mathrm{i}\kappa_{2}\alpha (\tilde{a}^{\dagger})^2\tilde{a})
            \nonumber\\&
            \hbar(-\mathrm{i}\kappa_{1}\tilde{a}\tilde{a}^{\dagger}-4\mathrm{i}\kappa_{2}|\alpha|^2\tilde{a}^{\dagger}\tilde{a}+\mathrm{i}\kappa_{2}(\alpha^{*})^{2}\tilde{a}^{2}-\mathrm{i}\kappa_{2}\alpha^{2}(\tilde{a}^{\dagger})^{2})
            \nonumber\\&            
            \hbar(-\mathrm{i}[\dot{\alpha}(t)\tilde{a}^{\dagger}-\dot{\alpha}^{*}(t)\tilde{a}]+\mathrm{i}\kappa_{1}\alpha \tilde{a}^{\dagger}-\mathrm{i}\kappa_{1}\alpha^{*}\tilde{a}+2\mathrm{i}\kappa_{2}\alpha (\alpha^{*})^2 \tilde{a}-2\mathrm{i}\kappa_{2}\alpha^{*}\alpha^2\tilde{a}^{\dagger})
      \ .
\end{align}
In terms of the effective Hamiltonian, we can write the master equation Eq.~\eqref{MasterEqRot} as follows
\begin{align}
      \label{EffHamMasterEqVDP}
            \dot{\rho}_{\alpha}(t)&=-\frac{\mathrm{i}}{\hbar}H^{(\alpha)}_{\text{eff}}\rho_{\alpha}(t)+\frac{\mathrm{i}}{\hbar}\rho_{\alpha}(t)(H^{(\alpha)}_{\text{eff}})^{*}+2\kappa_{2}\tilde{a}^{2}\rho_{\alpha} (\tilde{a}^{\dagger})^{2}+4\kappa_{2}\alpha^{*}\tilde{a}^{2}\rho_{\alpha} \tilde{a}^{\dagger}+4\kappa_{2}\alpha \tilde{a}\rho_{\alpha} (\tilde{a}^{\dagger})^{2}
            \nonumber\\&
            +2\kappa_{1}\tilde{a}^{\dagger}\rho_{\alpha} \tilde{a}+8\kappa_{2}|\alpha|^2\tilde{a}\rho_{\alpha}\tilde{a}^{\dagger}
        \ .
\end{align}
Now we have all the necessary elements in order to study the quantum fluctuations. First, we eliminate the linear terms in the Hamiltonian. This is achieved as long as the condition
\begin{equation}
      \label{QuantEqMotion}
            \dot{\alpha}(t)=\kappa_{1}\alpha(t)-2\kappa_{2}\alpha(t)|\alpha(t)|^2
\end{equation}
is satisfied. Eq.\eqref{QuantEqMotion} corresponds to the equation of motion of a Van der Pol oscillator.
In order to perform a semiclassical treatment of the master equation~\cite{SICarmichael,SIClerk}, we observe that $|\alpha(t)|\gg 1$. This is equivalent to saying that the action of the semiclassical system is greater than the action quanta, which allows us to perform a semiclassical approximation to the master equation up to quadratic order in the quantum fluctuations 
\begin{equation}
      \label{SemiclassMaster}
        \dot{\rho}_{\alpha}(t)\approx-\frac{\mathrm{i}}{\hbar}[\hat{H}_{\text{Q}}^{(\alpha)}(t),\rho_{\alpha}(t)]+2\kappa_{1}\left(\tilde{a}^{\dagger}\rho_{\alpha} \tilde{a}-\frac{1}{2}\{\rho_{\alpha},\tilde{a} \tilde{a}^{\dagger}\}\right)+8\kappa_{2}|\alpha|^2\left(\tilde{a}\rho_{\alpha}\tilde{a}^{\dagger}-\frac{1}{2}\{\rho_{\alpha},\tilde{a}^{\dagger}\tilde{a}\} \right)
      \ ,
\end{equation}
where
\begin{equation}
      \label{VDPQuantumFluct}
            \hat{H}_{\text{Q}}^{(\alpha)}(t)=\mathrm{i}\hbar\kappa_{2}[\alpha^{*}(t)]^{2}\tilde{a}^{2}-\mathrm{i}\hbar\kappa_{2}[\alpha(t)]^{2}(\tilde{a}^{\dagger})^{2}
      \ .
\end{equation}
\section{Initial conditions}
In the case of the uncoupled system $V=0$, a single Van der Pol oscillator~\cite{SISadeghpour,SIBruder} exhibits a limit cycle with radius $r_{0}=\sqrt{\frac{\kappa_{1}}{2\kappa_{2}}}=1.58$ for the parameters $\kappa_{2}=0.2\kappa_{1}$ used in the main paper. For convenience, in the coupled system we consider initial conditions at $t=0$ in such a way that each oscillator has the same amplitude $|\alpha_{l}(0)|\approx1.58$.
In addition, we consider phases obtained from a Gaussian distribution
$\phi_{l}(0)=\frac{\theta}{\sqrt{2\pi}\sigma}\exp[-\frac{(l-\mu)^2}{2\sigma^2}]$, where $-24\pi<\theta<24\pi$ is a random number, $\mu=N/2$ and $\sigma=9$. Fig.~\ref{Fig1SI}a) shows the initial conditions. In terms of the coordinates $\alpha_{l}(t)=\frac{Q_{l}(t)+\mathrm{i}P_{l}(t)}{\sqrt{2\hbar}}$, the initial conditions must satisfy $\sqrt{Q^{2}_{l}(0)+P^{2}_{l}(0)}\approx 2.24$, which defines the green circle in Fig.~\ref{Fig1SI}b).

In the main text, although we consider different coupling strengths $V$, we use the same initial conditions as in Fig.~\ref{Fig1SI} to obtain the chimera, synchronized and desynchronized states.
In order to obtain the snapshot of the chimera state depicted in Fig.~\ref{Fig3}a), we let the system evolve up to a time $\kappa_{1}t_{0}=3000.5$ for a coupling strength $V=1.2$. Correspondingly, to obtain the snapshot of the synchronized solution shown in Fig.~\ref{Fig3}b), we let the system evolve a time
$\kappa_{1}t_{\text{Syn}}=25.5$ for $V=1.6$. Finally, the snapshot of the desynchronized in Fig.~\ref{Fig3}c) is obtained after a time evolution $\kappa_{1}t_{\text{desyn}}=8000.5$ for $V=0.8$.

\begin{figure}
\includegraphics[width=0.60\textwidth,clip=true]{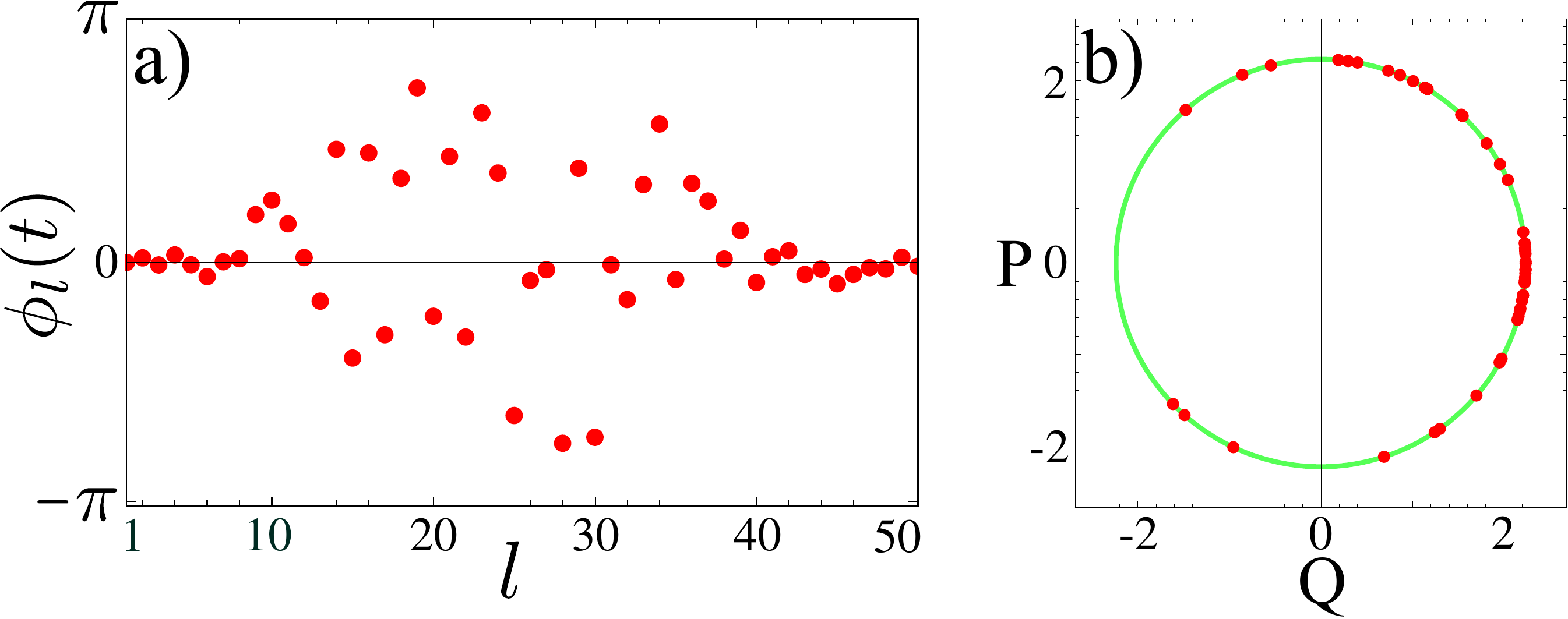}
    \caption{
    \label{Fig1SI}
   Initial conditions used in the main paper. a) Initial distribution of the phases $\phi_{l}(0)$ drawn randomly from a Gaussian distribution in space.
   b) Phase-space representation of the initial conditions for the oscillators. The green circle represents the limit cycle with radius $|\alpha_{l}(0)|\approx1.58$, where $\alpha_{l}(t)=\frac{Q_{l}(t)+\mathrm{i}P_{l}(t)}{\sqrt{2\hbar}}$. 
   Parameters $\hbar=1$, $d=10$, $\kappa_2=0.2\kappa_1$, and $N=50$.}
\end{figure}

\section{Explicit form of the Fokker-Planck equation}
Now let us define the Wigner representation of the density operator $\rho_{\bm{\alpha}}(t)$~\cite{SICarmichael}, as follows
\begin{equation}
      \label{WignerRepresentation}
            W_{\bm{\alpha}}(\bm{\tilde{z}})=\int \frac{d^{2N}\bm{\lambda}}{\pi^{2N}} e^{-\bm{\lambda}\cdot \bm{\tilde{z}}^{\ast}+\bm{\lambda}^{\ast}\cdot \bm{\tilde{z}}}\text{tr}\left[\rho_{\bm{\alpha}}(t)e^{-\bm{\lambda}\cdot \hat{\bm{\tilde{a}}}^{\dagger}+\bm{\lambda}^{\ast}\cdot \hat{\bm{\tilde{a}}}}\right]
            \ ,
\end{equation}
where $\bm{\lambda}=(\lambda_{1},\ldots,\lambda_{N})$ denote the integration variables, $\bm{\tilde{z}}=(\tilde{z}_{1},\ldots,\tilde{z}_{N})$ and $\tilde{z}_{l}=(\tilde{q}_{l}+\mathrm{i}\tilde{p}_{l})/\sqrt{2\hbar}$. In addition, the variables $\tilde{q}_{l}$ and $\tilde{p}_{l}$ denote position and momentum, respectively. 
The Wigner function is related to the Husimi function in the co-moving frame $Q_{\bm{\alpha}}(\bm{\tilde{z}})=\frac{1}{\pi}\bra{\bm{\tilde{z}}}\rho_{\bm{\alpha}}(t)\ket{\bm{\tilde{z}}}$ via the transformation \cite{SICarmichael}
\begin{equation}
      \label{WignerHusimi}
            Q_{\bm{\alpha}}(\bm{\tilde{z}})=\frac{2}{\pi}\int W_{\bm{\alpha}}(\bm{\tilde{x}})e^{-2|\bm{\tilde{z}}-\bm{\tilde{x}}|^2}d^{2N}\bm{\tilde{x}} 
      \ .
\end{equation}

By using the Wigner representation, the master equation Eq.~\eqref{SemiclassMasterNetwork} can be mapped into a Fokker-Planck equation~\cite{SICarmichael}
\begin{align}
      \label{WignerFokkerPlanck}
            \frac{\partial W_{\bm{\alpha}}}{\partial t} &=\sum_{l=1}^{N}\left[2\kappa_{2}(\alpha^{\ast}_{l})^{2}\partial_{\tilde{z}^{\ast}_{l}}\tilde{z}_{l}
            +(4\kappa_2|\alpha_{l}|^{2}-\kappa_{1})\partial_{\tilde{z}_{l}}\tilde{z}_{l}+\left(2\kappa_2|\alpha_{l}|^{2}+\frac{\kappa_{1}}{2}\right)\partial^2_{\tilde{z}_{l},\tilde{z}^{\ast}_{l}}\right]W_{\bm{\alpha}}
            \nonumber \\&
            -\mathrm{i}\frac{ V}{2d}\sum_{l=1}^{N}\sum_{\substack{m=l-d\\m\neq l}}^{l+d}(\partial_{\tilde{z}^{\ast}_{m}}\tilde{z}^{\ast}_{l}-\partial_{\tilde{z}_{l}}\tilde{z}_{l})W_{\bm{\alpha}}+
            \text{H.c}
            \ .
\end{align}
By defining $\bm{\tilde{R}}^{T}=(\tilde{q}_{1},\tilde{p}_{1},\dots,\tilde{q}_{N},\tilde{p}_{N})$,
the Fokker-Planck equation can be written also in terms of the quadratures $\tilde{q}_{l}$ and $\tilde{p}_{1}$
\begin{equation}
      \label{FokkerPlanckCanonical}
          \frac{\partial W_{\bm{\alpha}}}{\partial t}=-\sum^{2N}_{i=1}\mathscr{A}_{ij}(t)\partial_{\tilde{R}_{i}}(\tilde{R}_{j}W_{\bm{\alpha}})+\frac{1}{2}\sum^{2N}_{i=1}\mathscr{B}_{ij}(t)\partial^{2}_{\tilde{R}_{i},\tilde{R}_{j}}W_{\bm{\alpha}}
      \ .
\end{equation}
Although this Fokker-Planck equation has time dependent coefficients, one can derive an exact solution~\cite{SICarmichael}
\begin{equation}
      \label{SolFokkerPlanck}
            W_{\bm{\alpha}}(\bm{\tilde{R}},t)=\frac{\exp\left(-\frac{1}{2}\bm{\tilde{R}}^{T}\cdot\mathscr{C}^{-1}\cdot\bm{\tilde{R}}\right)}{(2\pi)^{N}\sqrt{\det \mathscr{C}}}
       \ ,
\end{equation}
where the covariance matrix $\mathscr{C}(t)$ is a solution of the differential equation $\dot{\mathscr{C}}(t)=\mathscr{A}(t)\mathscr{C}(t)+\mathscr{C}(t)\mathscr{A}^{T}(t)+\mathscr{B}(t)$.

%
\end{widetext}

\end{document}